\documentstyle[prb,aps]{revtex}

\begin{document}
\twocolumn[\hsize\textwidth\columnwidth\hsize\csname %
@twocolumnfalse\endcsname
\draft
\preprint{ver. 2.2} 
\title{Interference of a first-order transition with the formation\\ of a
spin-Peierls state in $\alpha^{\prime}$-NaV$_2$O$_5$?}
\author{M.\,K\"{o}ppen, D.\,Pankert, and R.\,Hauptmann}
\address{Institute of Solid State Physics, SFB 252, Darmstadt University
of Technology, \\
D-64289 Darmstadt, Germany}
\author{M.\,Lang, M.\,Weiden, C.\,Geibel, and F.\,Steglich}
\address{Max-Planck-Institute of Chemical Physics of Solids, \\
D-01187 Dresden, Germany}
\date{\today}
\maketitle

\begin{abstract}
We present results of high-resolution thermal-expansion and specific-heat
measurements on single crystalline $\alpha^{\prime}$-NaV$_2$O$_5$ .
We find clear evidence for two almost degenerate phase transitions associated with the 
formation of the dimerized state around 33\,K:
A sharp first-order transition at T$_1=(33\pm 0.1)$\,K slightly below the
onset of a second-order transition at T$_2^{onset}\approx(34\pm 0.1$)\,K.
The latter is accompanied by pronounced spontaneous strains.
Our results are consistent with a structural transformation at T$_1$ induced by
the incipient spin-Peierls (SP) order parameter above T$_2$=T$_{\rm SP}$.
\end{abstract}


]
\narrowtext

\section{Introduction}

The discovery of a spin-Peierls (SP) transition in the inorganic compound
CuGeO$_3$ by Hase {\em et al.}~\cite{Has93} has triggered a large amount of
investigations concerning the physical characteristics of the SP phase
transition. A detailed analysis of the magnetic susceptibility showed that
CuGeO$_3$ actually behaves differently from an ideal SP system. 
A significant next-nearest-neighbor coupling has to be invoked to understand
the magnetic properties of this compound\cite{Cas95}.\\ 
Recently, Isobe and Ueda\cite{Iso96} reported susceptibility results on 
another inorganic
system, polycrystalline $\alpha ^{\prime }$-NaV$_2$O$_5$, which shows  a 
similar rapid decrease in the magnetic susceptibility below
34\,K. This feature together with the existence of a broad maximum in the 
susceptibility 
around 350\,K which closely follows the prediction for an 
S=$\frac 12$ antiferromagnetic Heisenberg chain 
led the authors to propose 
$\alpha ^{\prime }$-NaV$%
_2$O$_5$ as a possible new SP compound.\\ 
In fact, the isotropic decrease of the magnetic susceptibility found for 
single 
crystalline material\cite{Wei97}, the occurrence of a lattice distortion 
below the transition
temperature as observed by X-ray\cite{Fuj97}, Raman-scattering\cite{Wei97},
thermal-expansion\cite{Wei97a} and NMR-measurements\cite{Oha97}
as well as the formation of a gap in the magnetic-excitation 
spectrum\cite{Iso96} are prominent characteristics of a spin-Peierls 
transition. In general,
for such a phenomenon to take place, the presence of spin-1/2 chains is 
considered a necessary condition.
From early structural investigations on polycrystalline
material a one-dimensional (1D) magnetic character was inferred \cite{Garpy},
implying two kinds of V chains running parallel to the b axis, one of which 
is formed by magnetic (S=1/2) V$^{4+}$, the other one by nonmagnetic
V$^{5+}$ ions. Such a charge-ordered state at T=300\,K, however, was found
to be incompatible with recent results of Raman-scattering 
experiments\cite{Lemm}. 
Moreover, a refined analysis of room temperature X-ray data indicate the 
existence of only one distinct V site\cite{Sch}, posing the question
whether a 1D magnetic structure is actually realized in this system.\\
So far, most of the work on $\alpha ^{\prime }$-NaV$_2$O$_5$ 
was aiming at an investigation  of the magnetic and structural aspects 
whereas
little is known about the thermodynamic properties. 
In this paper we will
present results of thermal-expansion and specific-heat measurements on an
$\alpha ^{\prime }$-NaV$_2$O$_5$ single crystal focussing on the phase 
transition  into the nonmagnetic low-T state.

\section{Experimental}

Single crystals of $\alpha^{\prime}$-NaV$_2$O$_5$ were grown by slowly
 cooling  a mixture of NaVO$_3$ 
and 
VO$_2$ with a molar ratio of 10:1 in an evacuated quartz tube. This 
assures that the melt is always Na rich - a condition necessary to achieve a 
Na content 
x=1,  the highest possible 
Na concentration in $\alpha^\prime$-Na$_x$V$_2$O$_5$\cite{Pou}. This is 
especially important as the magnetic properties of $\alpha^\prime$-Na$_x$V$_2$O$_5$
are strongly x dependent: already a slight Na deficiency removes 
the SP transition completely\cite{Wei97}.
The sample was cooled down from 800$^\circ$C with a cooling rate of 1\,K/h. 
The single crystals grown by this method could be easily separated from the 
NaVO$_3$-flux by 
boiling the obtained sample in water since NaVO$_3$ only is soluble in hot 
water.
The susceptibility of the single crystal (with outer dimensions 
a x b x c = 1.17 x 5.25 x 0.29mm$^3$ and mass m=5.41mg) selected for the 
present investigations shows 
a sharp transition at T$_{\rm SP} \approx$ 33\,K, accompanied by a
rapid and isotropic decrease in $\chi$(T) towards lower temperatures 
(not shown). The small 
upturn in $\chi$(T) for T$\to$0 is ascribed to the presence of impurities (or 
defects). Assuming their effective spin to be S=1/2, their concentration
is estimated to be $<$0.05\%, reflecting the high 
quality of the crystal.\\

The thermal expansion was measured by means of an ultra-high-resolution 
capacitance dilatometer\cite{Pot83} with a maximum sensitivity of $\Delta$%
l/l=10$^{-10}$. Length changes $\Delta$l(T)=l(T)-l(4.2\,K) were
detected employing a heating rate $\frac{d {\rm T}}{d {\rm t}}\le2.5$\,K/h.
The coefficient of thermal expansion $\alpha (T) = \frac{1}{l(T)}\frac{dl}{dT}$
 was approximated by the differential
quotient $\alpha (T) \approx \frac{\Delta l(T_2) -\Delta l(T_1)}{l(300{\rm %
K)\cdot (T_2 - T_1)}}$ with T=(T$_1$+T$_2$)/2. Measurements were performed
along the three principal axes. The specific heat was measured using an
AC-calorimeter\cite{Sul68}. A resistance heater is fed by
an AC-power with frequency $f$/2 causing the temperature of the sample to
oscillate with $f$. Within an appropriate range of frequencies,
the amplitude of the temperature oscillation, T$_{AC}$, is inversely proportional to the
heat capacity, C(T), of the sample. T$_{AC}$ was determined using a lock-in
amplifier. 
The calorimeter was calibrated by measuring a sample of pure copper.

\section{Thermal Expansion}

\label{TE} Figure \ref{ges}  gives a survey of the linear
thermal-expansion coefficient for T$\le$200\thinspace K measured along the
three principal axes. The data reveal a pronounced anisotropy 
consistent with the results of an X-ray 
study\cite{Iso96}. While  the
c-axis expansivity is extraordinarily large with a broad maximum around 175\,K,
 the b- ('in-chain direction')
and a-axis expansion coefficients are very small for temperatures
40\thinspace K$\le $T$\le$250\thinspace K. 
A blow-up of the data (cf.\ inset fig.\,\ref{ges}) shows that
$\alpha_a$ and $\alpha _b$ are negative up to about 150\,K yielding a broad 
minimum around 85\thinspace K and 75\,K, respectively. The low-temperature 
$\alpha$(T) behavior is 
dominated by
sharp phase-transition anomalies around the spin-Peierls transition 
temperature T$_{SP}\approx $\,33\thinspace
K. Figures \ref{alphasp} a-c give details of 
$\alpha $(T) near T$_{\rm SP}$ along the three principal axes (note 
the different
scales). 
Along the a- and c-axis we find large $\lambda$-type phase-transition anomalies
of comparable magnitude but opposite signs. A striking feature of
both data sets is the 
steepening of the slope of $\alpha$(T) on the low-T side of the 
transition, i.~e. slightly below
the temperature of the $\alpha$(T) extremum. As becomes more evident from the 
b-axis expansivity data (fig.~\ref{alphasp}c) we assign this behavior to 
the combined effects of two
phase transitions, which are almost degenerate in temperature (see discussion 
in section \ref{Dis}):
An extremely sharp phase-transition anomaly (at T$_1$) located right at the 
middle of a somewhat broadened one (at T$_2$). While the two corresponding 
phase-transition
anomalies are positive for $\alpha_a$ and negative for $\alpha_c$, they have 
opposite signs for $\alpha _b$.
Moreover, a careful inspection of the $\alpha(T)$ data along the three axes 
shows
that the sharp peak is actually split up into two very sharp structures 
separated by only 200\,mK. This is exemplarily shown for the b-axis data in the
inset of fig.~\ref{alphasp}. To exclude 
the possibility that this fine structure is caused by
inhomogeneities in the Na content, further experiments on an independently
prepared single crystal are in preparation.\\
The influence of a magnetic field of B=8\,T applied parallel to the b-axis on 
$\alpha_b$ is shown in fig.\,\ref{B}. Surprisingly, the field does not lead
to a homogenous shift of the two transitions towards lower temperatures. 
Rather it acts quite differently
on both sides of the transition at T$_1$: While the high-temperature flank of the
second-order transition (as e.\,g.\ , its onset temperature T$_2^{onset}$)
shows a comparatively strong reduction of $\Delta T_2^{onset}=-(0.36\pm0.1)$\,K,
no significant field dependence is found for $\alpha_b$ for temperatures 
below 
T$_1$, i.\,e.\ for T$\le$32.2\,K. Compared to the effect on T$_2^{onset}$, T$_1$
shows a much weaker field dependence of only 
$\Delta$T$_1$=\mbox{-(0.13$\pm$0.1)\,K.}\\
In order to clarify the nature of the two transitions and to 
separate their respective contributions, we plot in figs.~\ref{dl} and
\ref{dl2} the relative length changes, $\Delta$l$_i$/l, for i = a, b and c on 
the same temperature scale. As becomes obvious from fig.~\ref{dl}, 
the ordering phenomenon below 33\,K is accompanied by the appearance of 
pronounced spontaneous 
strains with large and partly 
compensating effects along the c- and a-axis and an only minor length change
along the b-axis. Besides this continuous lattice distortion, a detailed 
inspection of the data close to the transition 
provides  evidence for the first-order character of the sharp structure in
$\alpha(T)$
(cf.\ fig.~\ref{dl2}).
This becomes most clear for the b-axis data, where upon lowering the 
temperature
a sudden elongation occurs within a rather small temperature interval of only
0.45\,K centered around T$_1=(33.0\pm0.1$)\,K. This first-order transition is 
superimposed on the continuous change in the slope of 
$\Delta l/l$ from positive ($\alpha _b> 0$) for T$<$32.7\,K to negative 
($\alpha_b < 0$) for T$>$33.25\,K, which corresponds to the somewhat
broadened second-order
transition. We stress that in subsequent runs both transitions (including the 
splitting of the transition at T$_1$) could be 
reproduced in detail without any significant hysteresis at
T$_1$ upon cooling and warming through the transition temperature.\\
By projecting the finite width of the transition at T$_1$ on 
the a- and c-axis
data (vertical lines in fig. \ref{dl2}), one can clearly identify 
first-order-like behavior, i.\,e.\ sudden length changes, in the $\Delta l/l$ 
data also for the latter two axes.\\
For the determination of T$_2$, the transition temperature of the second-order
phase transition, we use the b-axis data where, due to their opposite signs, 
both 
transitions can be most clearly separated from each other. By simply 
subtracting in fig.\  \ref{alphasp}c the sharp
anomaly at T$_1$ from the $\alpha _b(T)$ data 
and replacing the remaining broadened transition by an idealized sharp one  
we find 
T$_2=(32.7\pm0.1)$\,K.
We note that a
non-vanishing contribution of the transition at T$_1$ to $\alpha (T)$ at 
lower
temperatures (neglected in the above procedure) might shift T$_2$ to slightly
higher values.\\
Our observation of pronounced spontaneous strains below T$_2$ indicate a substantial
coupling between magnetic and lattice degrees of freedom for the present
system. Within a Landau theory this coupling can be treated by adding
to the Gibbs free energy  
an interaction term 
\begin{equation}\label{coup}
G_{int}=
\sum_{mn}\xi_{i,m,n}\epsilon _i^mQ^n\hspace{.5cm}(m,n\in {\cal N)},
\end{equation}
 where
$\xi _{i,m,n}$ are the coupling constants between the order parameter
$Q$ and the strain elements $\epsilon _i$ along the $i$th direction. In
general, within the constraints given by the symmetry of the system, 
different coupling types
(n,m $>$0) may be possible. For a linear-quadratic strain-order-parameter 
coupling, $\xi_i\epsilon_iQ^2$, as e.\,g.~, realized in CuGeO$_3$\cite{Wink}, the 
condition of a 
stress-free system then yields $\epsilon_i(T)\propto Q^2(T)$, i.\,e.\ the
spontaneous
strains vary as the square of the order parameter. To extract the
$\epsilon _i$ (i=a,b,c),  associated with the spin-Peierls transition, for 
the present system one has
to subtract from the $\Delta l_i/l$ data (cf.\ fig.~\ref{dl}) the
non-relevant background contribution. To this end we employed a smooth 
extrapolation  of the $\Delta l/l$ data from T$>$T$_{\rm 2}$ into the low-T
phase. Although, due to the rather high transition temperature substantial
uncertainties have to be invoked for the background contribution, an only
minor effect is found as for the conclusion drawn below.
The resulting spontaneous strains normalized to their respective values at
4\,K are plotted in fig.~\ref{OP} as a function of reduced temperature 
$\frac{T_2-T}{T_2}$ on logarithmic scales. For all three axes we find 
power-law behavior, $\epsilon _i \propto \left(\frac{T_2-T}{T_2}\right)^\gamma$,
 over almost one decade in 
temperature. Deviations at low temperatures are likely to be caused by 
uncertainties in the background contribution. The tendency towards smaller
exponents visible for temperatures close to the transition is assigned to
the influence of one-dimensional fluctuations. From the slopes of the linear
parts in fig.\,\ref{OP} the exponent $\gamma$ can be determined. Most surprisingly,
the $\gamma$-values for the a-, b- and c-axis strains of respectively 
(0.23$\pm $0.05), (0.7$\pm 0.15$) and (0.36$\pm $0.1) differ significantly from
each other.
In particular $\gamma_b$, the exponent for the 'in-chain' strain is strongly 
enhanced - the same tendency, although much less pronounced  was found also 
for CuGeO$_3$\cite{Wink}.
As for the origin of this behavior one may think of (i) an anisotropy in the
coupling constants $\xi _{i,m,n}$ which causes different exponents in the
leading terms of eqn.~\ref{coup} along the various axes or (ii) a finite 
coupling to the order parameter
of the first-order transition. Since for the b-axis, the
$\alpha(T)$ anomalies associated with the respective transitions at T$_1$ and T$_2$ have opposite
signs, the one at T$_2$ being comparatively small, a rather strong influence of
the lattice distortion at T$_1$ on 
 $\epsilon_b$ appears plausible.\\

A further observation made in the present thermal-expansion
study is worth mentioning:  A comparison of the volume-expansion coefficient, $\beta$, as determined
from the present measurements on single crystalline material, $\beta _{sc}=
\alpha_a +\alpha_b +\alpha_c$, with that derived from polycrystalline samples,
$\beta_{pc}=3\cdot\alpha_{pc}$, reveals striking discrepancies (see ref.
\onlinecite{Wei97a} for $\alpha _{pc}$). Although T$_{\rm SP}$ is almost 
identical
for the two variants, the absolute value as well as the temperature variation 
of $\beta$ is strikingly different both at and above T$_{\rm SP}$. A similar 
observation was already made 
for CuGeO$_3$ (see ref.\ \onlinecite{Weid} and \onlinecite{Wink} for poly-
and single-crystalline samples, respectively). Up to now we
have no explanation for this discrepancy. Texture effects in the polycrystal
used in ref.\,\onlinecite{Weid} at the origin of this problem can be safely excluded\cite{Koe1}.

\section{Specific Heat}

\label{SH} 
\begin{figure}
\caption{Specific heat, C(T), as C {\em vs} T (a) and C(T)/T {\em vs}
T (b). Width of shaded area in (b) indicates the temperature range of the 
first-order transition as determined from $\alpha(T)$ measurements.}
\label{sph}
\end{figure}
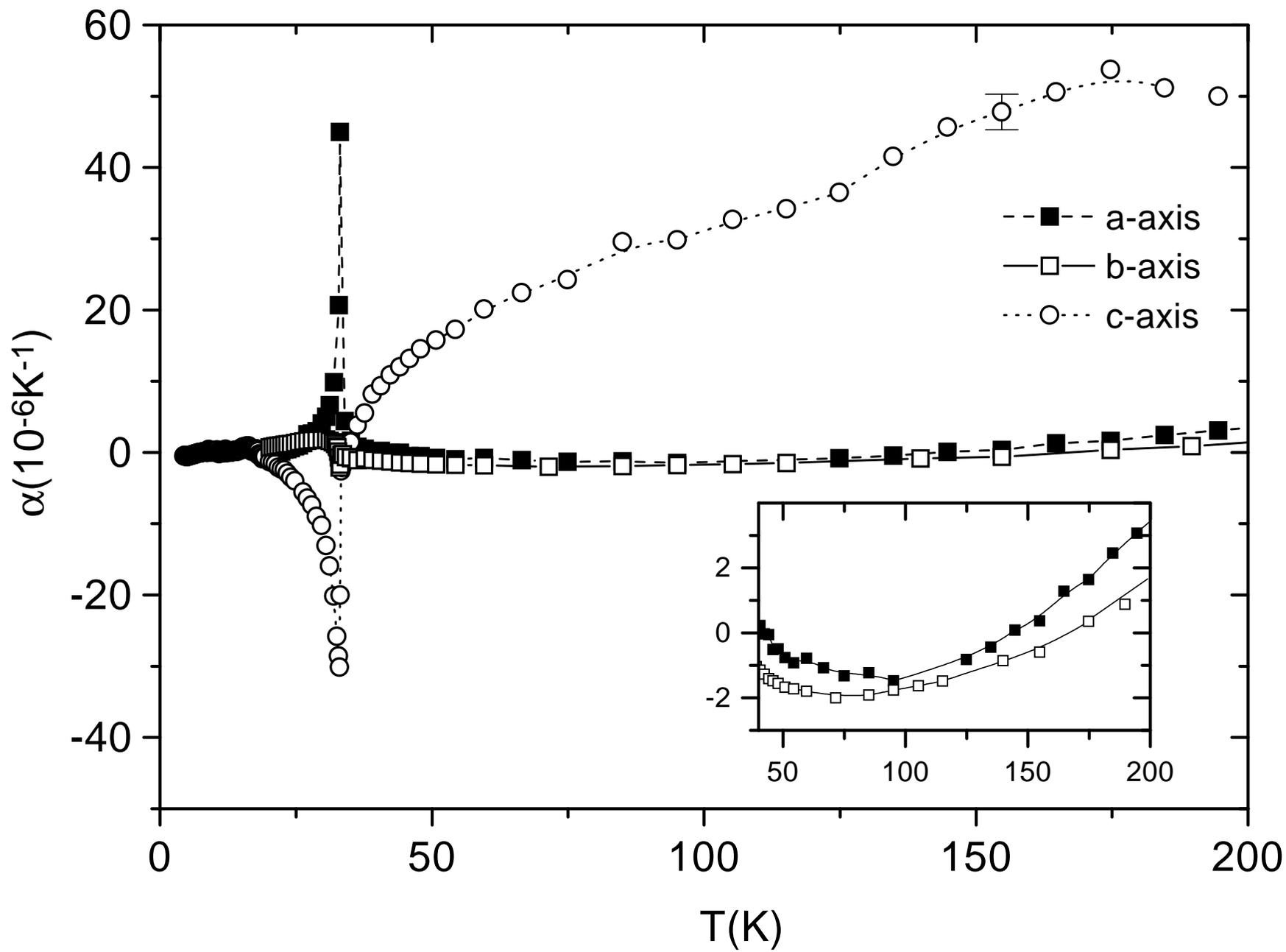
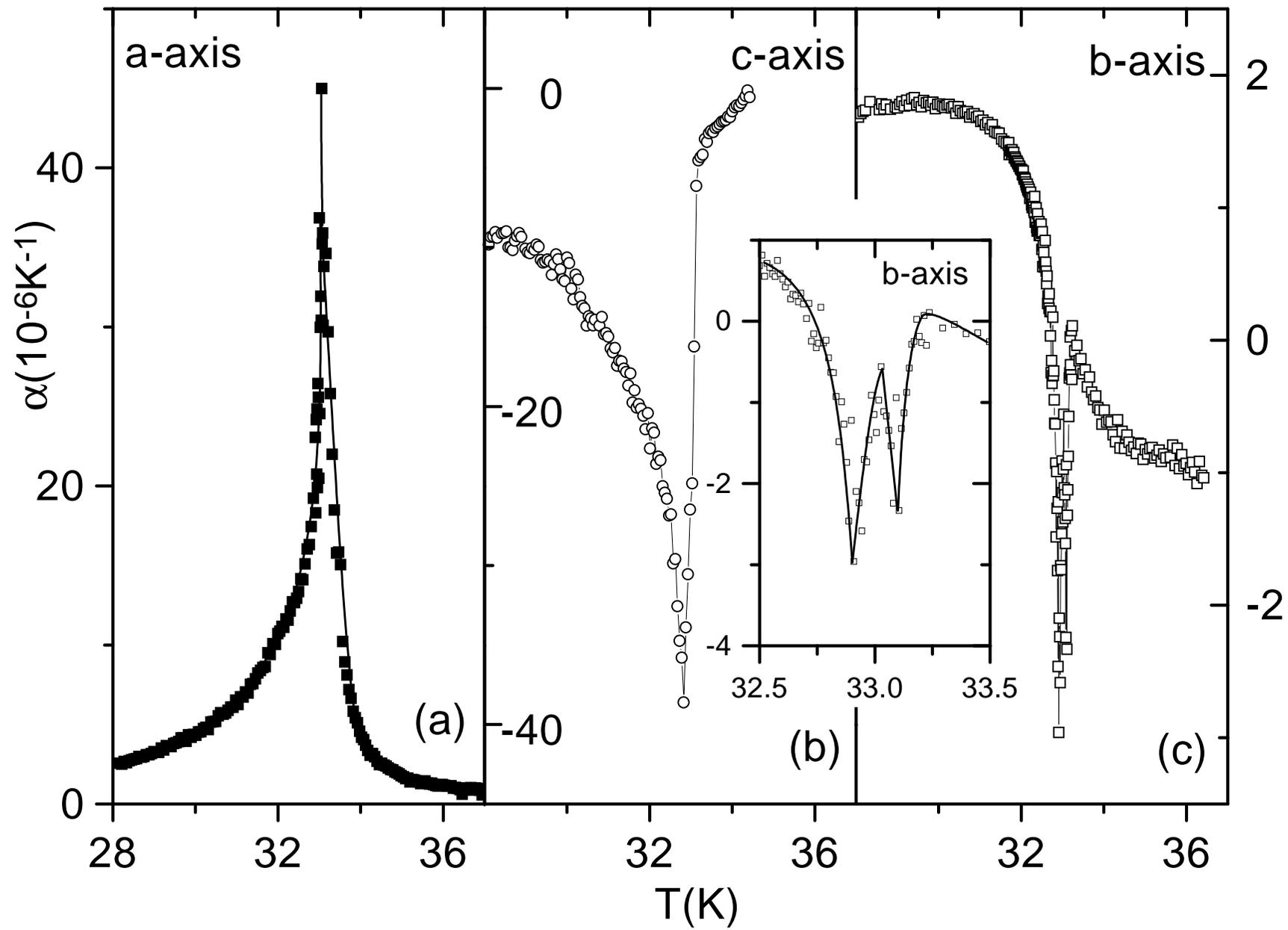
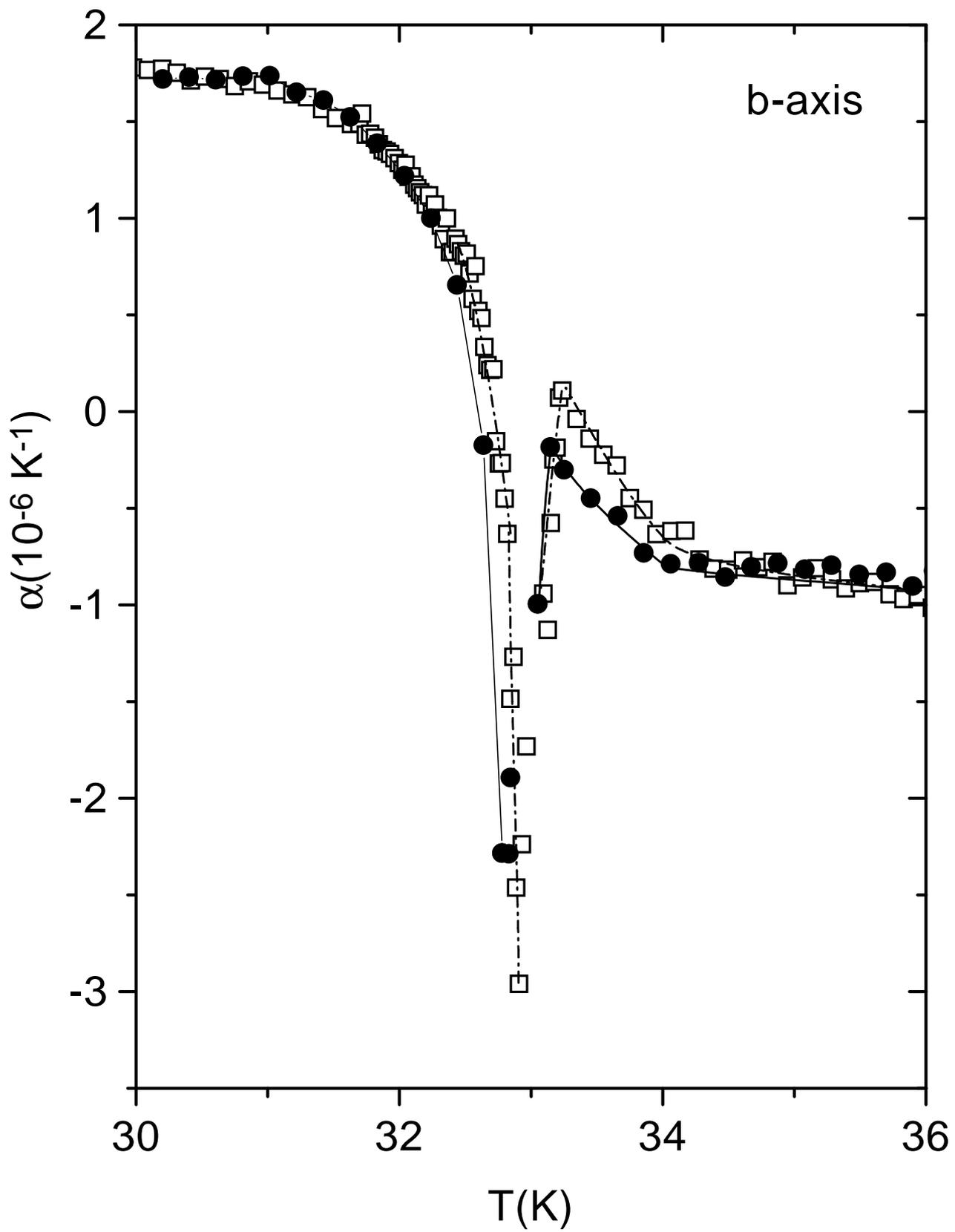
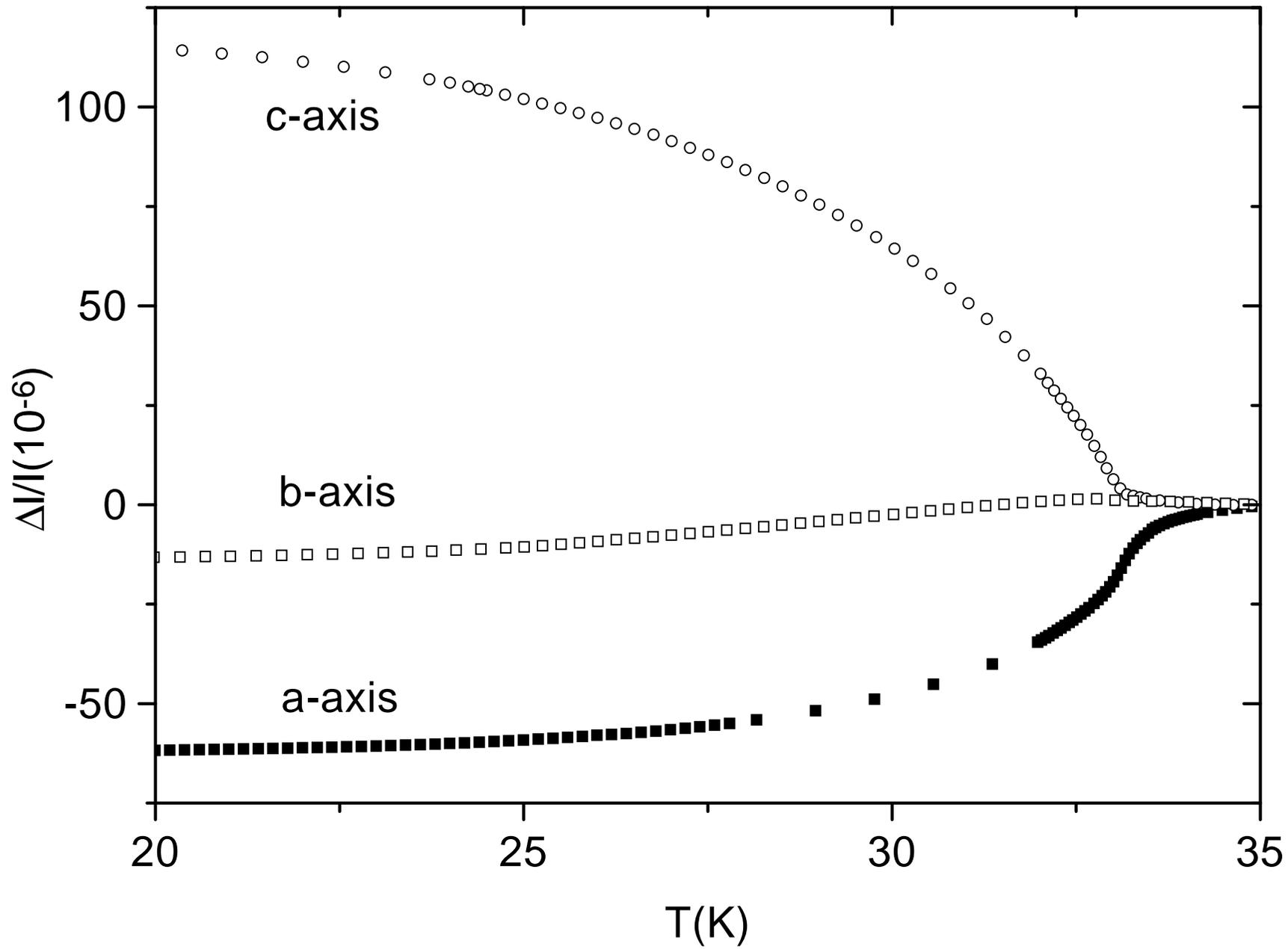
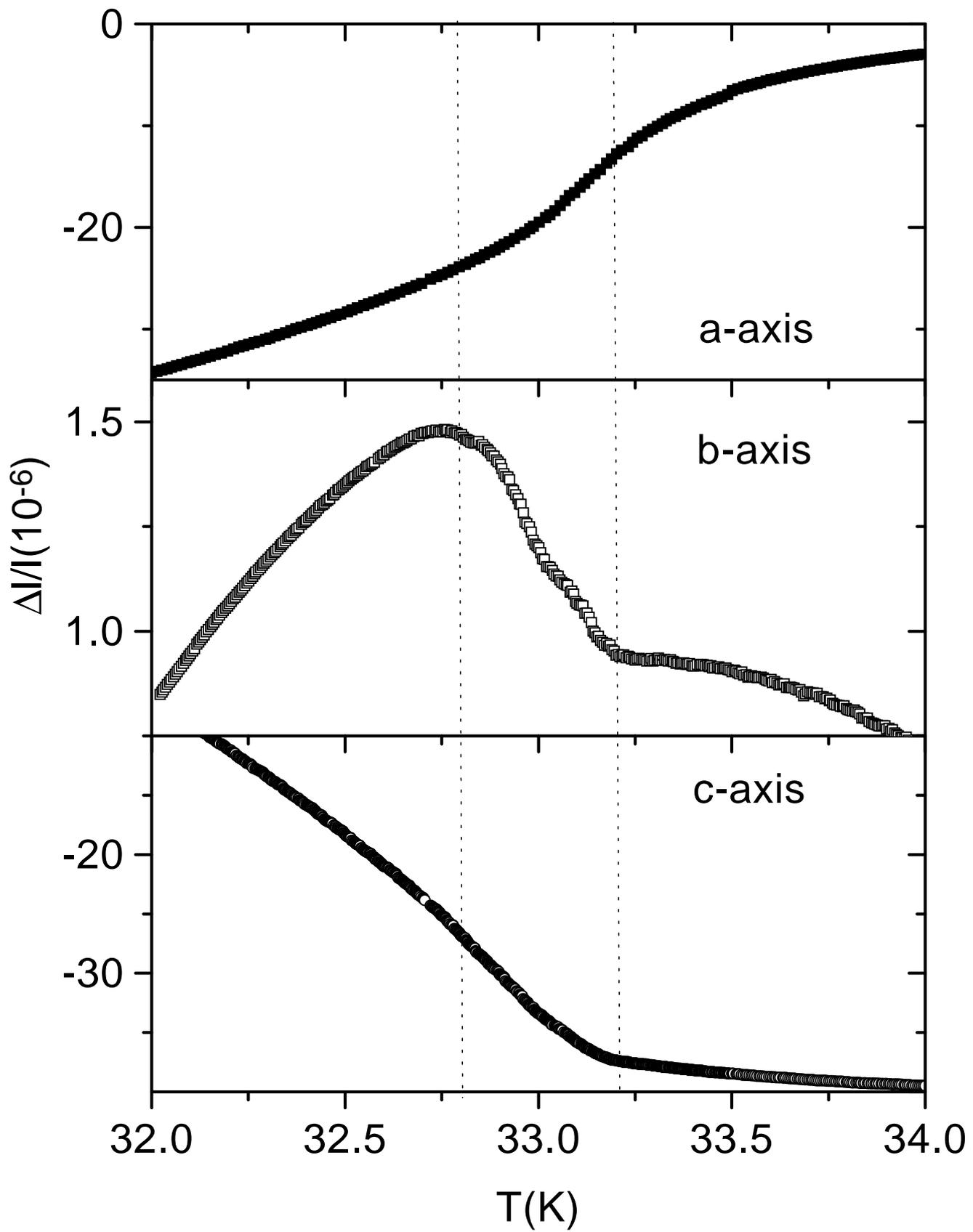
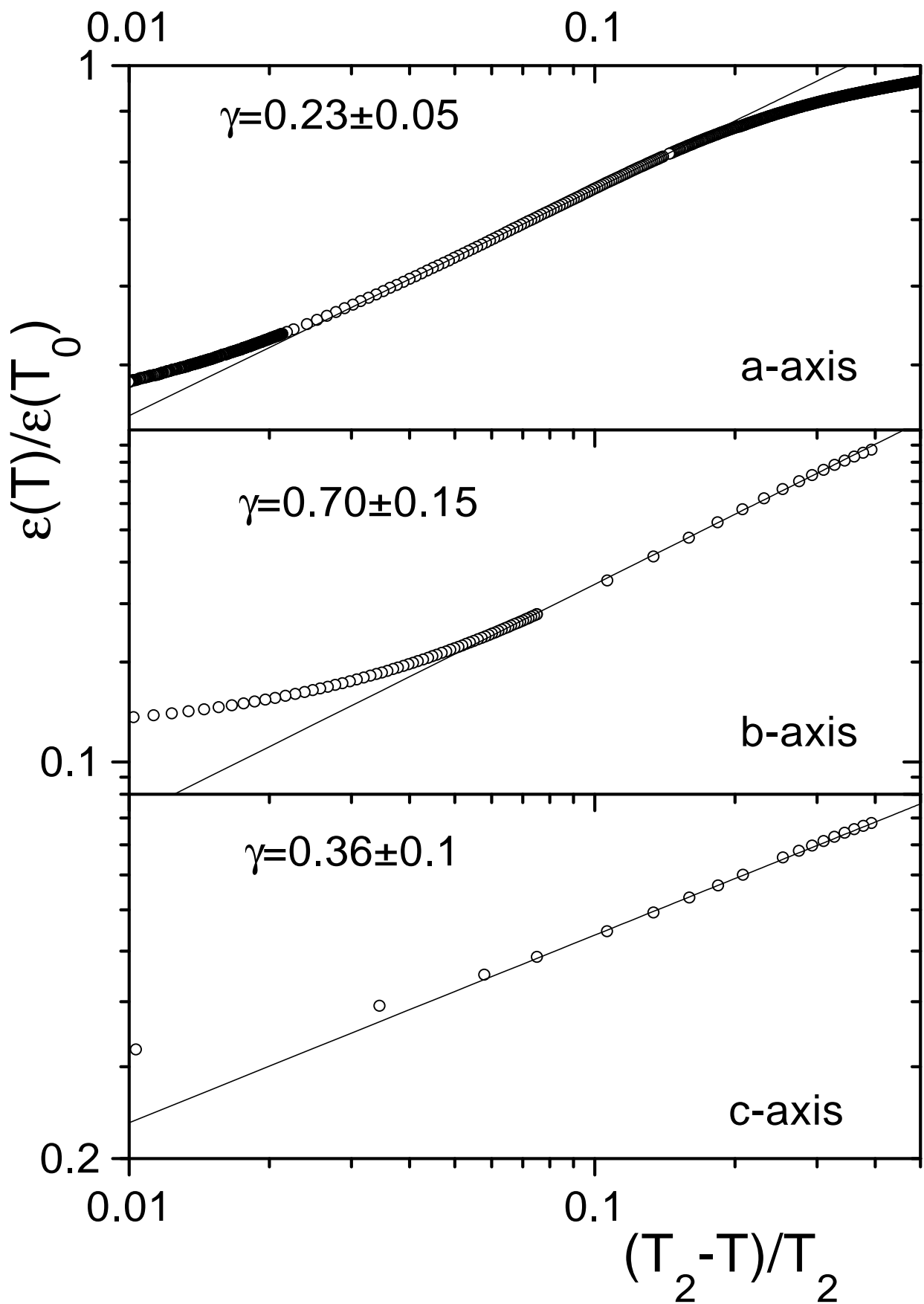
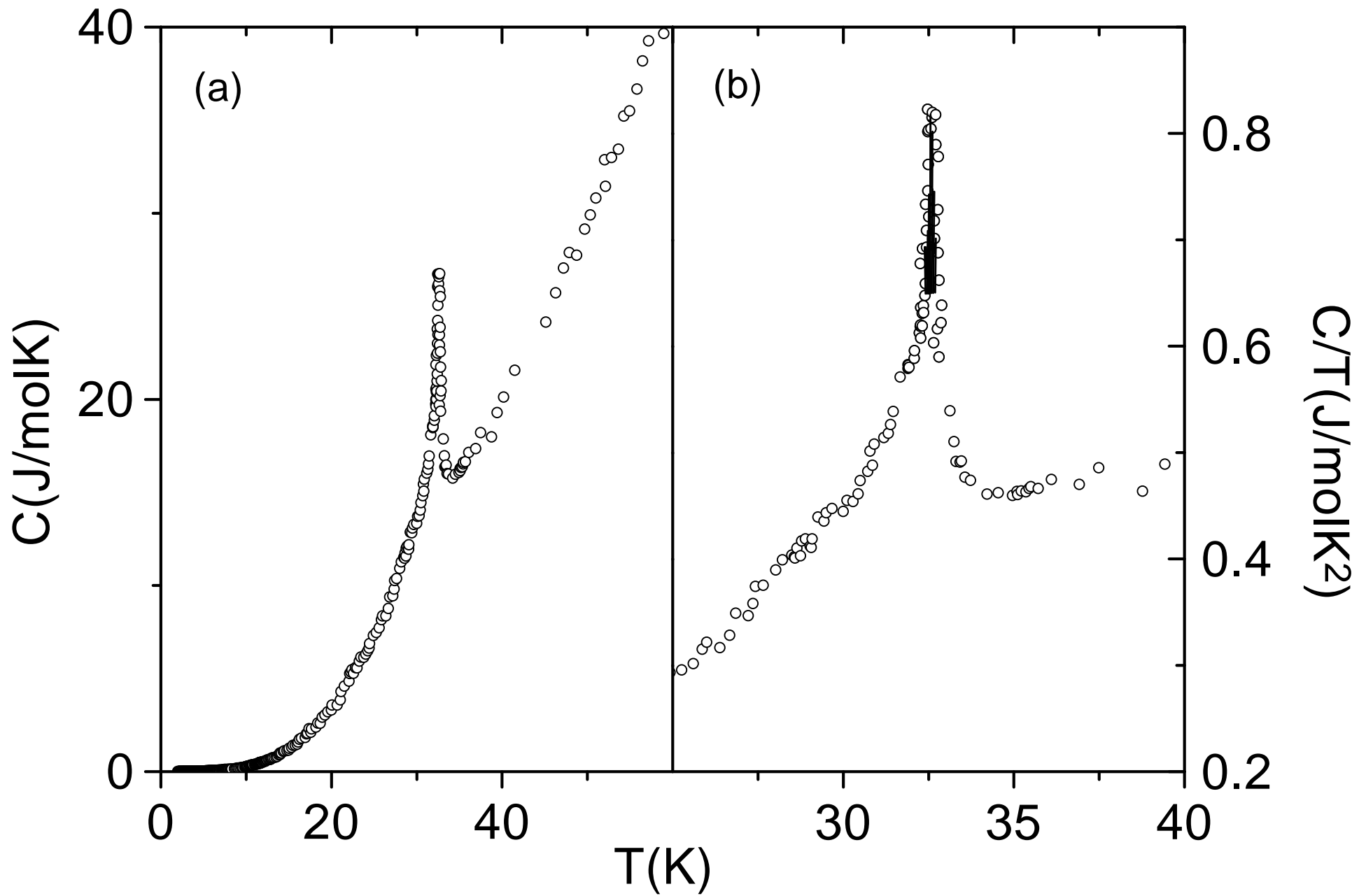

In order to give a full thermodynamical analysis of the two 
coexisting phase
transitions, we have performed supplementary measurements of the specific 
heat, C(T), on
the same single crystal in the temperature range 1.8K$\le T\le 60$\,K. 
Fig.~\ref{sph}a shows the specific heat as C(T) {\em vs} T over the whole
temperature range investigated.
The SP transition manifests itself in the sharp maximum around 33\,K.
An excellent fit of the low-temperature data, i.\,e. T$\le$12\,K, is provided
by adding to the phonon term 
C$_{\rm ph}=\beta \cdot T^3$ a magnetic contribution 
C$_{\rm mag}=a \exp (-\Delta /T)$, which corresponds to  singlet-triplet
excitations across an energy gap $\Delta$.
For the zero-temperature value we find $\Delta = (84 \pm 
10)$\,K, in good agreement with the results from measurements of the magnetic
susceptibility (85$\pm$15\,K)\cite{Weid}, the nuclear 
magnetic resonance
(98\,K)\cite{Oha97}, as well as the electron-spin resonance 
(92$\pm 20$\,K)\cite{Vasi} and
(85$\pm$20\,K)\cite{Pal}. These values are slightly smaller than 
$\Delta$=114\,K derived from inelastic 
neutron-scattering experiments\cite{Fuj97}.
Fig.~\ref{sph}b shows details of C(T) around 33\,K as  C(T)/T 
{\em vs} T.
With the knowledge of $\alpha$(T) (fig.~\ref{alphasp})  one might also 
separate two different 
contributions to C(T): A sharp maximum associated with the first-order
transition at T$_1$ (width of shaded area in fig.~\ref{sph}b is taken from 
the anomaly in $\alpha_b(T)$  at
T$_1$, cf.~fig.\,\ref{alphasp}c) superimposed on a somewhat broadened 
second-order transition at T$_2$. 
We stress that as a consequence of the AC technique employed here implying
temperature oscillations of the sample of  
$\pm50$\,mK at T=33\,K,
sharp structures become necessarily broadened in our C(T) data
compared to $\alpha$(T).

\section{Discussion}
\label{Dis}
The results presented in the preceding paragraphs suggest the existence of two
closely spaced phase transitions associated with the formation of a 
dimerized state in $\alpha ^{\prime }$-NaV$_2$O$_5$ around 33\,K: A sharp 
first-order transition at T$_1$=(33.0$\pm$0.1)\,K slightly above a continuous phase
transition at T$_2$=(32.7$\pm$0.1)\,K.\\
As an alternative interpretation one might think of the possibility that the observed features
reflect a {\em single} phase transition which is of weakly first order. Since the
b-axis $\alpha(T)$ data around the transition temperature reveal two 
anomalies of opposite signs, this scenario implies that the sharp but small
negative peak in $\alpha_b$ {\em must} be a 'secondary effect' caused - via 
the
three-dimensional elastic couplings - by the much larger effects in the 
orthogonal
directions. As will be discussed below, however, this interpretation appears 
incompatible with the field dependence of $\alpha_b$ and is thus discarded.\\

The occurrence of spontaneous strains below T$_2$ suggests that the latter
 transition
is driven by the strain-order-parameter coupling resulting in the formation
of a gap in the spin-excitation spectrum, i.~e.\ T$_2$=T$_{\rm SP}$. 
Then the questions 
at hand are: (i) what is the
nature of the first-order transition and (ii) what is its role in the formation
of the spin-Peierls ground state? As for the nature of the transition at T$_1$
it is tempting to assign this effect to a structural transformation.
The substantial magneto-elastic coupling, as highlighted by the pronounced 
spontaneous strains below T$_{\rm SP}$, then naturally implies a coupling of
 both order parameters.\\
   As a possible scenario we 
propose that, upon lowering the temperature, the incipient 
spin-Peierls order parameter
induces via strain-order-parameter coupling a structural transition
at T$_1$. The latter then stabilizes the spin-Peierls transition.
Note that along the a- and c-axis, where the dominant effect on the lattice 
is observed upon cooling below 33\,K, the length changes associated with the 
transition at T$_1$, 
$\Delta l|_{T_1}$, proceed along the same direction as $\epsilon_a$ and
$\epsilon_c$, respectively (cf.~fig.\,\ref{dl2}).
The above assignment of the two transitions is also consistent with the 
response  in $\alpha_b$ when a
magnetic field of 8\,T is applied along the b-axis (cf.\ fig.\,\ref{B}): While 
the
onset of the transition at T$_2$ shows a significant reduction of 
$\Delta T_2^{onset}$=\mbox{-0.36\,K,} T$_1$ is only slightly reduced in this 
field by 
$\Delta$T$_1$=\mbox{-0.13\,K.} Even more striking is the fact that the low-T 
end of 
the continuous transition at T$_2$, i.~e.\ $\alpha_b$(T) for T$<$T$_1$, does 
not show any significant
field dependence at all.
These shifts have to be compared with the field-induced reduction, 
$\Delta T_{\rm SP}^{cal}$, expected for an ideal spin-Peierls transition 
\cite{Bul} and actually found for all known spin-Peierls systems\cite{Bray,Has93}.
Employing the results of a Hartree-Fock approximation 
\begin{equation}
\frac{{\rm T_{SP}^{cal}(H)-T_{SP}(0)}}{{\rm T_{SP}(0)}}=
-a(\frac{g_b\mu _BH}{2k_B\cdot {\rm T_{SP}(0)}})^2
\end{equation}
where g$_b$=1.972 is the g-factor\cite{Vasi}, $\mu _B$ the Bohr magneton,
k$_B$ the Boltzmann constant and a=0.44\cite{Bul}, we find 
$\Delta T_{\rm SP}^{cal}$=-0.375\,K. From the excellent agreement with the
experimentally observed reduction of T$_2^{onset}$ we infer that the high-T
side of the transition, i.~e.\ at T$>$T$_1$, reflects the behavior of a common
spin-Peierls transition. 
The much weaker field dependence of T$_1$ and the almost field-independent
behavior below T$_1$ rule out the possibility that the sharp feature in
$\alpha_b$ at T$_1$ is a secondary effect caused by the pronounced anomalies in
the orthogonal directions. In this case a homogenous shift of the two features
at T$_1$ and T$_2$ in a magnetic field would be expected. Rather it indicates
 the action of another interfering mechanism.\\
A direct proof of our assignment of two phase transitions would be 
provided by pressure experiments,
where uniaxial strain is applied along the b-axis. From the different signs of
the phase-transition anomalies $\Delta \alpha_b$ and $\Delta l_b$ associated
with the respective transitions at T$_1$ and T$_2$, a pressure-induced 
separation of the transitions can be expected: Applying the Ehrenfest-relation
and the Clausius-Clapeyron equation to the second- and first-order transition,
respectively, we predict a positive pressure dependence of the
transition at T$_2$ 
($\left(\frac{\partial {\rm T_{2}}}{\partial {\rm p_b}}\right)_{p_b \to 0}>0$)
and a negative pressure dependence for the first-order
transition ($\left(\frac{\partial {\rm T_{1}}}{\partial {\rm p_b}}\right)_{p_b \to 0}<0$). \\

In conclusion, thermodynamical investigations on 
$\alpha^{\prime}$-NaV$_2$O$_5$ reveal indications for two almost degenerate 
phase transitions 
around 33\,K associated with the formation of a nonmagnetic ground state.
The second-order transition at T$_2$=(32.7$\pm$0.1)\,K shows the signatures
of a SP transition as to the occurence of spontaneous strains and the field
dependence of its onset temperature. In this sense we assign the first-order
transition at T$_1$=(33$\pm$0.1)\,K to a structural transformation induced by 
the incipient SP order parameter at T$>$T$_2$. In general, for this phenomenon to take
place, the existence of spin-1/2 chains is considered a necessary condition.
In light of the results of recent structural investigations at room temperature
which are inconsistent with the above requirements a detailed T-dependent 
determination of the magnetic structure appears badly needed.

\subsection*{Acknowledgment}

This work was supported by the BMBF grant No.\ 13N6608/1.

\begin{figure}
\caption{Coefficient of thermal expansion, $\alpha$, {\em vs}
T along the three orthorhombic crystal axes. Inset: blow up of 
$\alpha$ {\em vs} T for
the a- and b-axis.}
\label{ges}
\end{figure}
\begin{figure}
\caption{Coefficient of thermal expansion, $\alpha$, {\em vs} T for
28\,K$<$T$<$37\,K along the various axes. 
Inset shows details of the negative peak in $\alpha_b$ on expanded scales.}
\label{alphasp}
\end{figure}
\begin{figure}
\caption{Coefficient  of thermal expansion, $\alpha_b$, for B=0 ($\Box $) 
and B=8\,T ($\bullet $), applied parallel to the b-axis. Lines are guides to the
eyes. Data points around the double structure near T$_1$ have been omitted 
for clarity.}
\label{B}
\end{figure}
\begin{figure}
\caption{Relative length changes $\Delta l/l$ {\em vs} T 
along the three crystal axes shifted so that the curves meet at 35\,K.}
\label{dl}
\end{figure}
\begin{figure}
\caption{Relative length change $\Delta l/l$ {\em vs} T on expanded scales
close to the SP transition. Vertical lines indicate the temperature range of
the first-order transition.}
\label{dl2}
\end{figure}
\begin{figure}
\caption{Normalized spontaneous strains {\em vs} reduced temperature
on logarithmic scales. Solid lines represent fits of the form 
$\epsilon \propto (1-T/T_2)^\gamma$.
}
\label{OP}
\end{figure}
\end{document}